\documentstyle[aps,multicol,pra,epsfig]{revtex}
\begin{document}
\draft

\title{Modulational and Parametric Instabilities of the Discrete Nonlinear 
Schr\"odinger Equation}
\author{Z. Rapti$^1$, P.G. Kevrekidis$^1$, A. Smerzi$^{2,3}$ 
and A.R. Bishop$^3$}
\address{
$^1$  Department of Mathematics and Statistics, University of
Massachusetts, Amherst, MA 01003-4515, USA \\
$^2$ Istituto Nazionale di Fisica per la Materia BEC-CRS and
Dipartimento di Fisica, Universita' di Trento, I-38050 Povo, Italy\\
$^3$ Theoretical Division and Center for Nonlinear Studies, Los Alamos
National Laboratory, Los Alamos, NM 87545, USA\\
}
\date{\today}
\maketitle
\begin{abstract}
We examine the modulational and parametric
instabilities arising in a non-autonomous, discrete nonlinear Schr{\"o}dinger
equation setting. The principal motivation for our study stems from
the dynamics of Bose-Einstein condensates trapped in a deep optical lattice.
We find that under periodic variations of the
heights of the interwell barriers (or equivalently of the scattering
length), additionally to the modulational instability,
a window of parametric instability becomes available to the system.
We explore this instability through multiple-scale analysis and
identify it numerically. Its principal dynamical characteristic is
that, typically, it develops over much larger times than the 
modulational instability, a feature that is qualitatively justified
by comparison of the corresponding instability growth rates.
\end{abstract}

\vspace{2mm}
\section{ Introduction}
The modulational instability (MI) is a general feature of
discrete as well as continuum nonlinear wave equations.
For this instability, a specific
range of wavenumbers of plane wave profiles of the form
$u(x,t) \sim \exp(i (k x - \omega t))$ becomes unstable
to modulations, leading to an exponential 
growth of the unstable modes and eventually to delocalization
(upon excitation of such wavenumbers) in momentum space. 
That is equivalent to localization in position space,
and hence the formation of localized, coherent solitary wave structures
\cite{sulem}.

The realizations of this instability 
span a diverse set of disciplines ranging
from fluid dynamics \cite{benjamin67} (where it is usually referred
to as the Benjamin-Feir instability) 
and nonlinear optics \cite{ostrovskii69}
to plasma physics \cite{taniuti68}.
One of the earliest contexts in which its significance
was appreciated was the linear stability analysis 
of deep water waves. 
It was much later recognized that the
conditions for MI would be significantly modified for discrete 
settings relevant to, for instance,
the local denaturation of DNA \cite{peyrard93} 
or coupled  arrays of optical waveguides 
\cite{dnc0,morandotti99}. 
In the latter case,
a relevant model is the discrete nonlinear Schr\"odinger equation (DNLS), 
and its MI conditions were discussed in \cite{kivshar92}. 
Most recently, the MI has been recognized as responsible for dephasing
and localization
phenomena in the context of Bose-Einstein condensates (BEC) in the
presence of an optical lattice i.e., a sinusoidal external 
potential \cite{wu01,sm02,konotop02,cat02}.

In the context of BECs which are among the principal motivations
of this work, another interesting possibility arises.
For a ``deep'' optical lattice (i.e., if
the wells of the spatially periodic potential 
are well-separated and sufficiently high), 
it has been shown that the relevant mean field model that
describes the behavior of the condensate, at $T=0$, is the discrete
nonlinear Schr{\"o}dinger (DNLS) equation \cite{sm02,sm01,konotop01,alfimov02}.
The optical lattice can be created by two counterpropagating laser beames
forming a standing wave interference pattern. 

Our interest in the present work is in introducing an explicit 
temporally periodic modulation in the coefficients of the DNLS
and examining the instabilities that may arise (for uniform solutions).
In the BEC setting, there is a number of potential realizations of
such a non-autonomous DNLS equation. For instance,
the heights of the interwell barriers of the optical lattice are 
proportional to the intensity of the lasers, and can be easily periodically
modulated in time. This induces an oscillating 
tunneling amplitude of the condensates between adjacent wells, as well as
an oscillating interaction energy of the condensates trapped in each well.
An alternative possibility involves the 
periodic modulation of the scattering length of the interaction
between the atoms 
via a Feshbach resonance, 
i.e., an external magnetic field; see e.g., \cite{inouye}.
The possibility of this, so-called, Feshbach resonance
management (FRM) of the interaction  has generated a large 
interest recently
due its robust effect on patterns, coherent structures and 
its potential for avoiding collapse, see e.g., 
\cite{frm,frm1,frm0,frm2,frm3}.

In this short communication, we revisit the modulational instability
criteria in the DNLS equation (which were originally derived in
\cite{kivshar92}), but in the presence of the periodic modulation
of the DNLS tunneling and interaction parameters, as motivated above.
We choose the simplest possible periodic modulation (a sinusoidal
variation of the atomic scattering length) and derive the modulational
stability equation, which in this case becomes a modified Mathieu equation.
In the absence of the periodic perturbation we recover the results
of \cite{kivshar92}. In the presence of such a term, an additional,
{\it parametric} instability becomes possible.
These new domains of instability appear due to parametric resonance, 
whenever the parameters of a system vary periodically with time.
In  contrast to ordinary resonance, where we have growth proportional to the 
time variable, here the growth is exponential, as in the case of 
customary modulational instability; see e.g., \cite{landau73}.
We implement a multiple-scale
expansion to identify the instability domain boundaries and subsequently
numerically examine our analytical predictions. The numerical investigations
indicate that this parametric resonance sets in over much longer time scales
than the modulational instability (for the same perturbation amplitude).
Similar considerations for the continuum problem can be found 
in \cite{frm,abd}.In 
\cite{frm} a slightly more restricted ansatz was used [$q=0$ was used in
our Eq. (\ref{deq5}) below], and in \cite{abd}
it was the dispersion that was varying with time, instead of 
the nonlinearity strength as in our case. The continuum limit of
our analytical results (modulo the relevant rescalings/adjustments) has 
been found to agree with the results of \cite{frm,abd}.

Our presentation is structured as follows. In section II we present
the mathematical framework and our analytical results. In section III we 
corroborate these results with numerical simulations. Finally, in section
IV, we summarize our results and present our conclusions.

\section{ Setup and Analytical Considerations}

We study the discrete nonlinear 
Schr\"{o}dinger equation in the form:

\begin{equation}
i \dot{\psi}_{n}=-D(t)(\psi_{n+1}+\psi_{n-1}-2 \psi_{n})+a(t) |\psi_{n}|^{2} 
\psi_{n}+E(t) \psi_{n},
\label{deq1}
\end{equation}
where the coefficients $D(t)$, $a(t)$ and $E(t)$ are time dependent.
If one sets 
$\psi_{n}(t)=\phi_{n}(t) e^{-i \int_{0}^{t}(2 D(s)+E(s)) ds}$, 
then (\ref{deq1}) is reduced to 

\begin{equation}
i \dot{\phi}_{n}=-D(t)(\phi_{n+1}+\phi_{n-1})+a(t) |\phi_{n}|^{2} \phi_{n}.
\label{deq2}
\end{equation}
Now, if we additionally set 
$\tau=\tau(t)$, $\tilde{\phi}_{n}(\tau)=\phi_{n}(t)$,
$\tilde{D}(\tau)=D(t)$, $\tilde{a}(\tau)=a(t)$ and choose 
$\frac{d \tau}{d t}=D(t)$,
then (\ref{deq2}) is equivalent to 

\begin{equation}
i \dot{\phi}_{n}=-(\phi_{n+1}+\phi_{n-1})+b(t) |\phi_{n}|^{2} \phi_{n},
\label{deq3}
\end{equation}  
where $b(t)=\frac{a(t)}{D(t)}$, and we must notice at this point that, with 
a small abuse of notation, we will continue to use $t$ instead of $\tau$.
From now on, we will study (\ref{deq3}), where we assume that 

\begin{equation}
b(t)=1+\epsilon \sin(\omega t),
\label{deq4}
\end{equation}
as motivated earlier.
It is easily verified that

\begin{equation}
v_{n}(t)=e^{i (2 \cos {q} \,t-\int_{0}^{t}b(s)ds+q n)},
\label{deq5}
\end{equation}
where $q$ is the wavenumber, is an exact solution of (\ref{deq3}). In order to 
examine the modulational stability of this plane wave solution, we use the ansatz

\begin{equation}
\phi_{n}(t)=v_{n}(t) \left[1+\tilde{\epsilon} \left(\alpha(t) e^{i k n}+\beta(t) e^{-i k n} \right)\right],
\label{deq6}
\end{equation}
where $k$ is the perturbation wavenumber and $\alpha(t)$, $\beta(t)$ are 
complex, time dependent fields.
Substituting (\ref{deq6}) into (\ref{deq3}) and keeping 
only the $O(\tilde{\epsilon})$
terms, we obtain the following first order, coupled system for $\alpha$ 
and $\beta^{*}$, where $^{*}$ denotes the complex conjugate:

\begin{eqnarray}
i \dot{\alpha}&=&4 \sin(\frac{2 q+k}{2}) \sin(\frac{k}{2}) \alpha+
b(\alpha+\beta^{*})
\\
i \dot{\beta^{*}}&=&4 \sin(\frac{2 q-k}{2}) \sin(\frac{k}{2}) \beta^{*}-
b(\alpha+\beta^{*}).
\end{eqnarray}
These can be combined to give a second order equation:

\begin{equation}
\ddot{\alpha}+(2 i B-\frac{\dot{b}}{b}) \dot{\alpha}+
(A^{2}-B^{2}+2 A b-i\frac{\dot{b}}{b} (A+B)) \alpha=0,
\label{deq7}
\end{equation}
where 

\begin{eqnarray}
A&=&2 \cos q (1-\cos k)
\label{deqA}
\\
B&=&2 \sin q \sin k
\end{eqnarray} 
are constants that depend 
only on the wavenumber $q$ and the perurbation wavenumber $k$.
It is worth noticing that if $\epsilon=0$ in (\ref{deq4}), then the 
modulational instability 
criterion reads $A (A+2)<0$ which is in accordance with Eq. ($10$) of 
\cite{kivshar92}.
Next, we substitute (\ref{deq4}) into (\ref{deq7}) to obtain, up to 
order $O(\epsilon)$, the equation:

\begin{equation}
\ddot{\alpha}+(2 i B-\epsilon \omega \cos(\omega t)) \dot{\alpha}+
(A^{2}-B^{2}+2 A+\epsilon (2 A \sin(\omega t)-i \omega (A+B) 
\cos(\omega t))\alpha=0.
\label{deq8}
\end{equation}
This is an equation with periodic coefficients \cite{magnus}, hence
it is natural to examine whether parametric instabilities may arise
due to the temporal modulation. 
In view of this perturbation, if we use 
a regular perurbation expansion of $\alpha$ as a 
series in $\epsilon$, 
$\alpha=\alpha_{0}+\epsilon \alpha_{1}+O(\epsilon^{2})$, we find that secular 
terms exist if $A=A_{0}=-1+\sqrt{1+\frac{\omega^{2}}{4}}$, which is equivalent
to $q=q_{0}=\arccos(\frac{A_{0}}{2 (1-\cos{k})})$. Thus, it is 
necessary to implement a multiple scale 
analysis \cite{bender}, expanding $q$ as a series in $\epsilon$, 
\begin{equation}
q=q_{0}+ \epsilon q_{1}+O(\epsilon^{2}). 
\label{deq9}
\end{equation}
After lengthy but straightforward calculations,
 it is found that the value of $q_{1}$ is 
\begin{equation}
q_{1}=\pm \frac{-1+ \sqrt{1+\frac{\omega^2}{4}}}
{4 \sqrt{1+\frac{\omega^2}{4}} (1-\cos{k}) \sin{q_{0}}}.
\label{deq10}
\end{equation}
After the substitution of (\ref{deq10}) into (\ref{deq9}) and using 
the expression (\ref{deqA}) for A, we obtain the boundaries of the 
instability domain on the $(q,\epsilon)$ plane to be:

\begin{equation}
\pm \epsilon=\frac{4 \sqrt{1+
\frac{\omega^2}{4}} (1-\cos{k}) \sin{q_{0}}}
{-1+\sqrt{1+\frac{\omega^2}{4}}} (q-q_{0}).
\label{deq11}
\end{equation}
\section{Numerical Results}
In order to check the validity of our analytical approach, we have 
performed numerical simulations of the equations of motion (\ref{deq3})
using a fourth order Runge-Kutta scheme. 
The parameters of the system have been chosen to be $\epsilon=0.01$, 
$\omega=1$ and $k=\pi/2$. The initial condition, in accordance
with Eq. (\ref{deq6}), is a modulated  wave
\begin{equation}
u_{n}=e^{i k n}+ \epsilon e^{i(q+k) n}.
\label{deqnew}
\end{equation}
The simulations have been performed with a chain of
$N=1380$ sites, with periodic boundary conditions so that the wave numbers 
$q(k)$ defined modulo $2 \pi$ in the lattice, are of the form
$q=2 \pi r/N$ ($k=2 \pi R/N$), where $r$ ($R$) is an integer (see also
\cite{kivshar92}).

\begin{figure}
\centering
{\epsfig{file=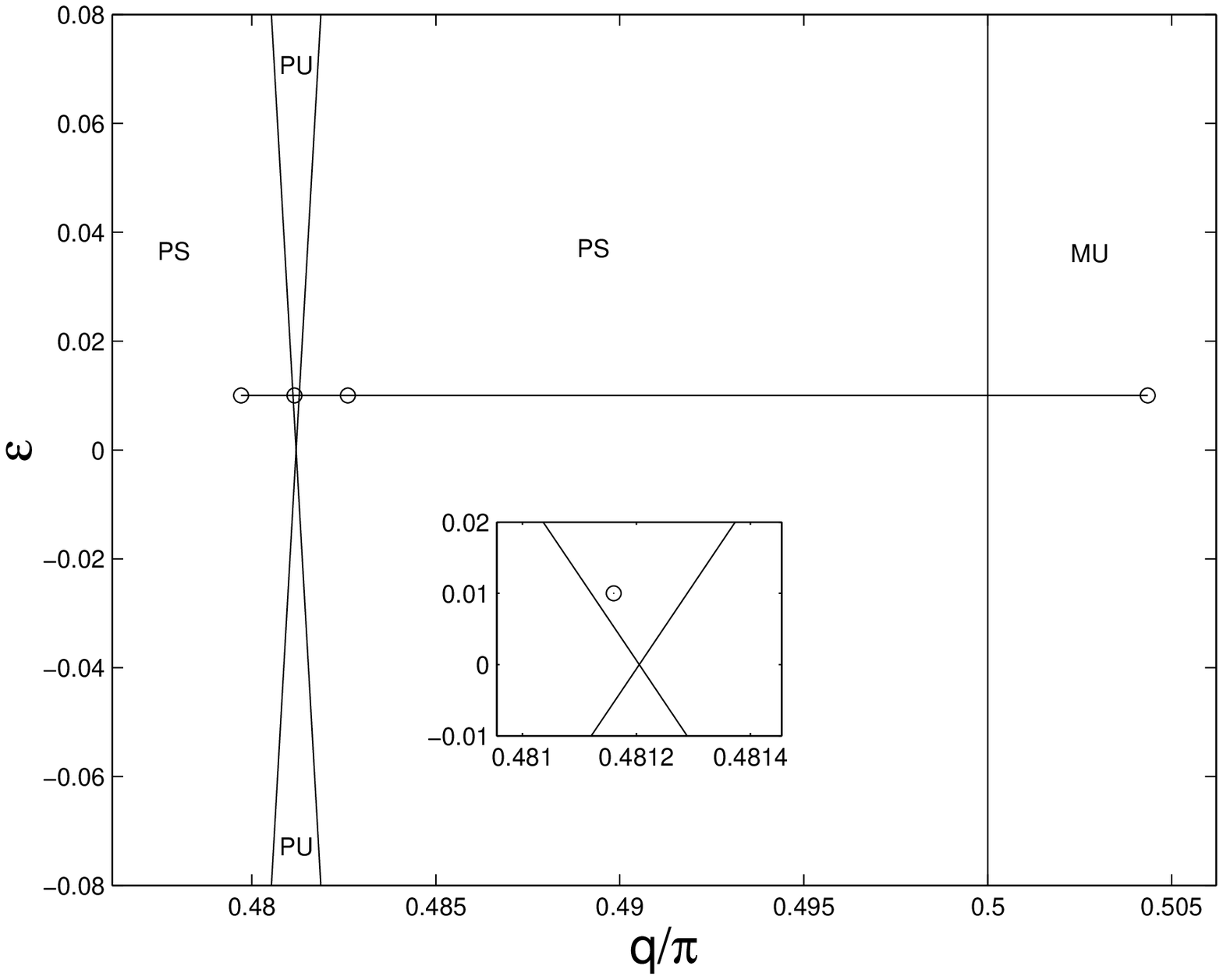, width=6.3cm,angle=0, clip=}}
\caption{The plane of ($\epsilon,q$) where the region of
the parametric instability represented by Eq. (\ref{deq11}) is
shown by two straight lines. Wavenumbers between the two
``stability limits'' will be (to leading order) parametrically unstable.
We also highlight 4 wavenumbers for which we conduct numerical simulations,
namely $q=1.50705$, $q=1.51161$, $q=1.51616$ and $q=1.58446$ , that 
lie on the line $\epsilon=0.01$ (see text).We use
PS (PU) to indicate the parametrically stable (unstable) regions, 
respectively, 
and MU to indicate modulationally unstable ones. The 
vertical
line $q=\pi/2$ is the boundary between the modulationally stable-unstable 
regions. 
}
\label{sfig0}
\end{figure}

Fig. \ref{sfig0} shows the windows of parametric instability represented
by the stability limits of Eq. (\ref{deq11}). For wavenumbers between the
two straight lines, parametric instability is theoretically predicted
from the results of Section II. To illustrate this point, we
select 4 wavenumbers, one of which ($q=1.50705$) is both modulationally
and parametrically stable; the second ($q=1.51161$) is parametrically
unstable; the third ($q=1.51616$) is again in the window of stability,
while the fourth one ($q=1.58446$) is modulationally unstable.

\begin{figure}
\centering
{\epsfig{file=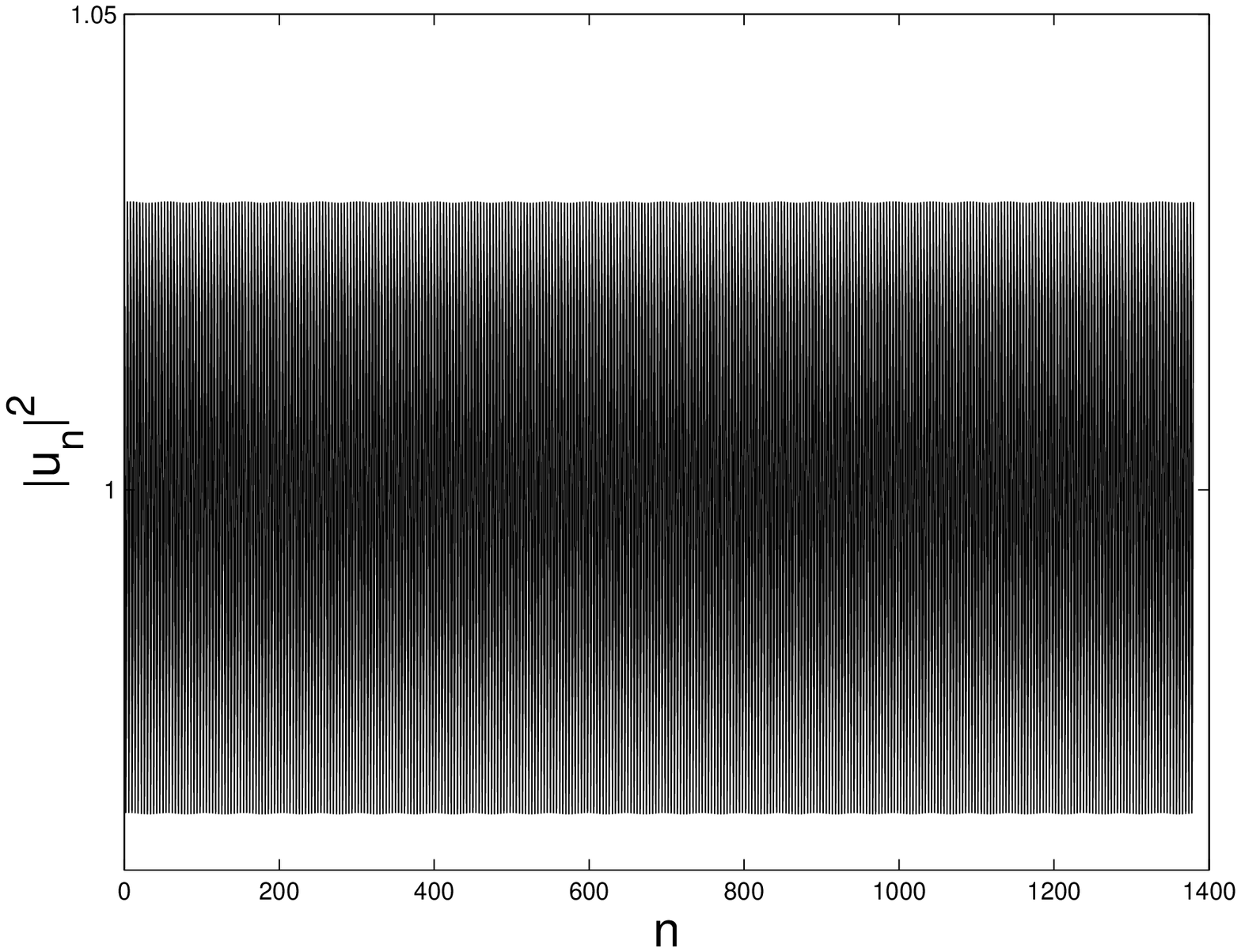, width=6.3cm,angle=0, clip=}}
{\epsfig{file=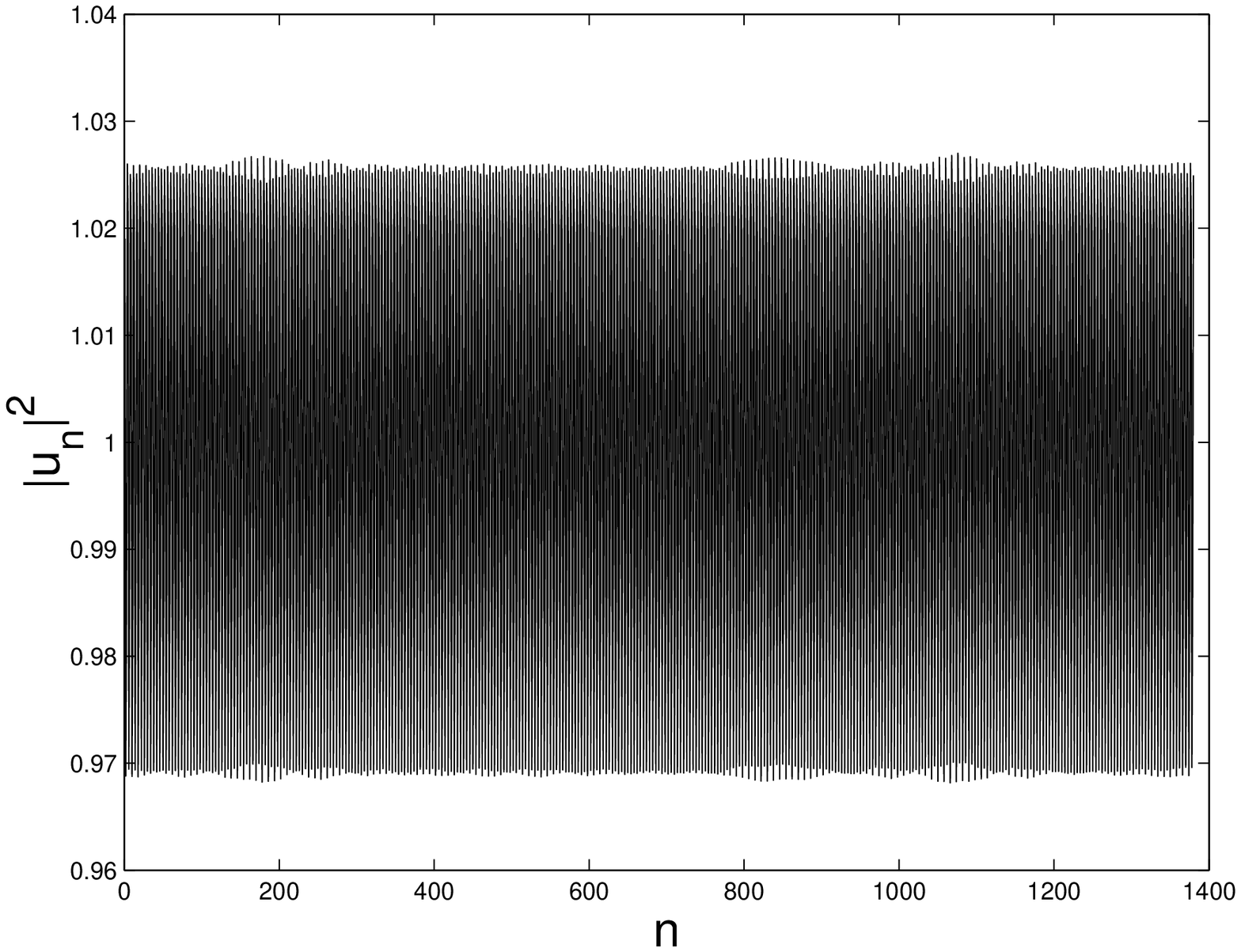, width=6.3cm,angle=0, clip=}}
\caption{Both panels show the case where we have modulational and 
parametric stability. The left corresponds to $q=1.50705.$ and $t=800$ and the 
right to $q=1.51616$ and $t=800$. The squared amplitude ($|u|^2$) of the
solution is shown in the panels as a function of the lattice site $n$.} 
\label{sfig1}
\end{figure}

In the left panel of Fig. \ref{sfig1}, we indeed observe at time 
$t=800$ that the perturbed wave is both parametrically and
modulationally stable. On the other hand, the right panel displays
$q=1.51616$, which should also be parametrically stable according to
our linear theory. However, for longer times than the ones displayed
in Fig. \ref{sfig1}, the latter wavenumber
has been observed to become unstable. 
We conjecture that this instability is due to a higher order 
parametric resonance, not captured within the theory of section II.

\begin{figure}
\centering
{\epsfig{file=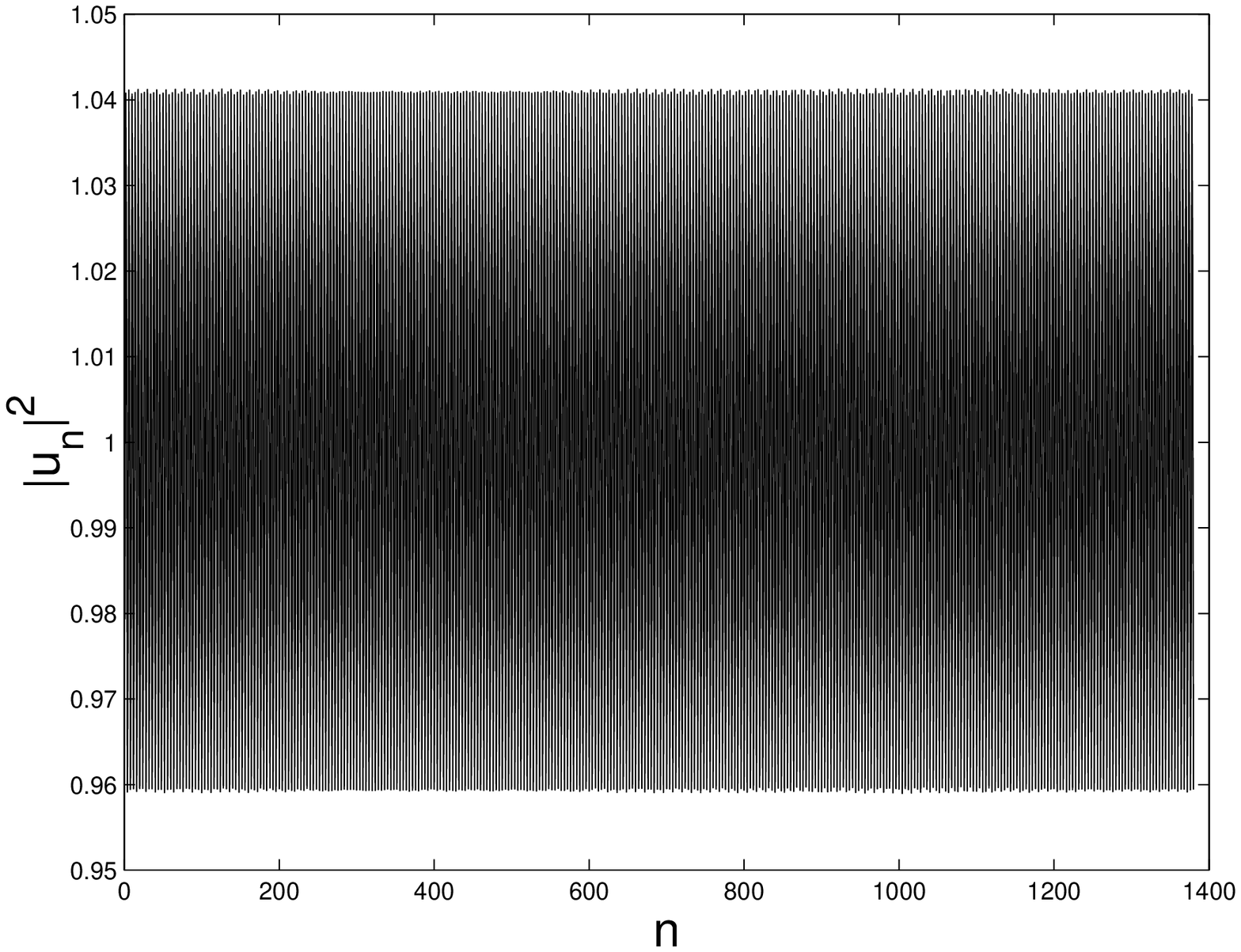, width=6.3cm,angle=0, clip=}}
{\epsfig{file=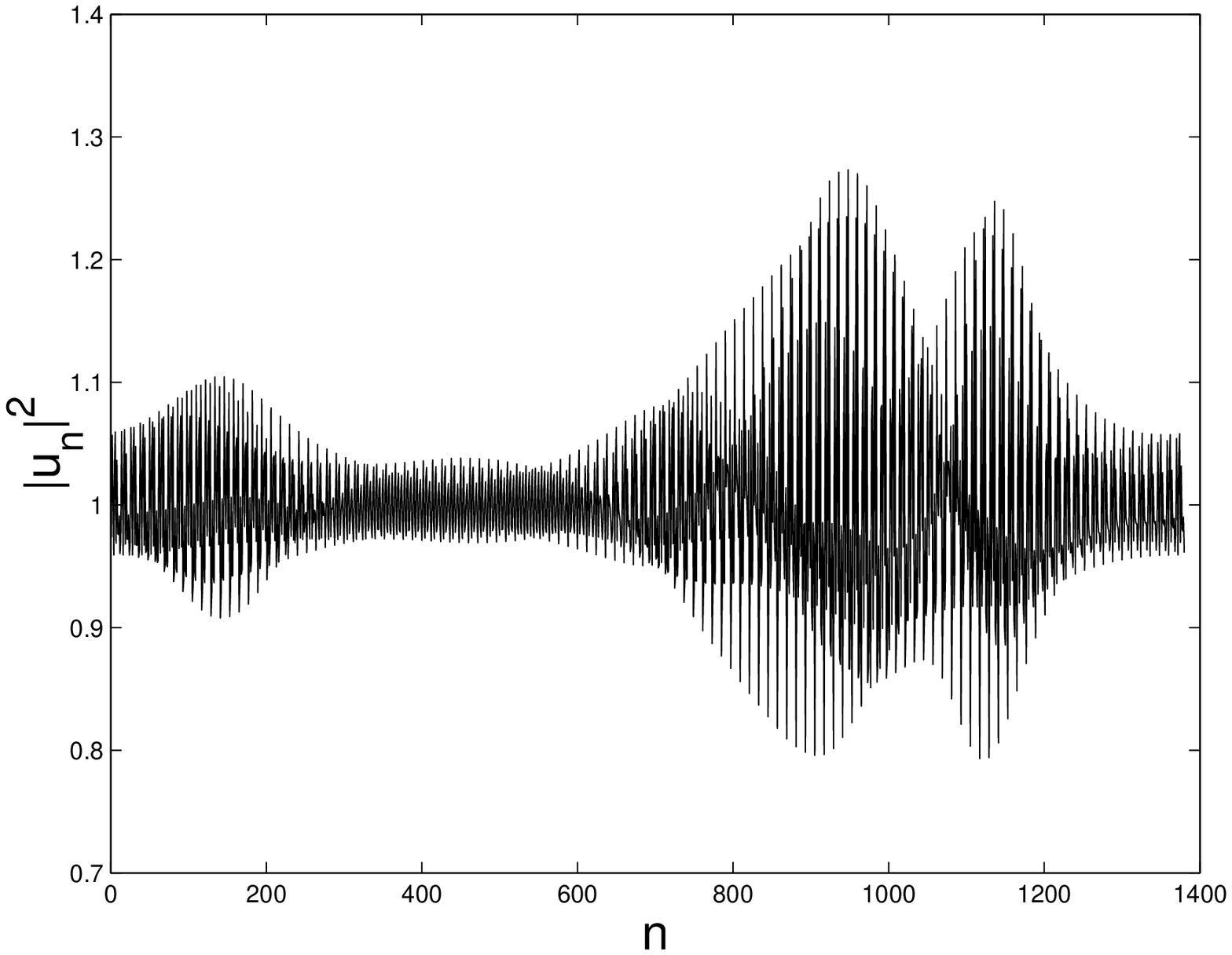, width=6.3cm,angle=0, clip=}}
\caption{The case which is modulationally stable, but  
parametrically unstable. The left panel shows the square amplitude 
of the field before 
the onset of the instability at $t=570$, while the right one after it, at 
$t=655$. In this case $q=1.51161$.}
\label{dfig2}
\end{figure}

The case of Fig. \ref{dfig2} displays the time evolution of the 
wavenumber predicted to be in the window of parametric instability
($q=1.511616$). We indeed find that at times $t=570$, the configuration
becomes parametrically unstable.

\begin{figure}
\centering
{\epsfig{file=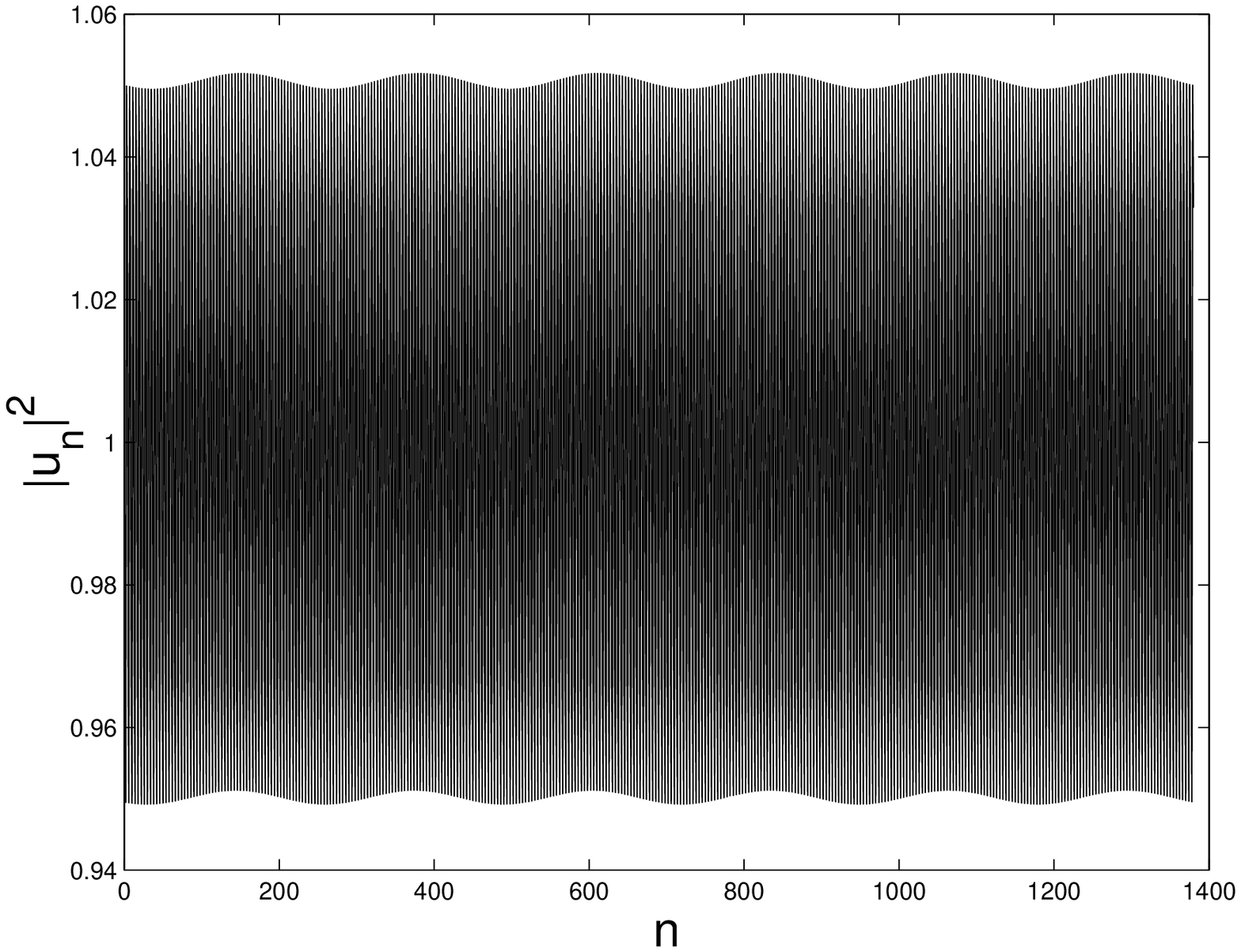, width=6.3cm,angle=0, clip=}}
{\epsfig{file=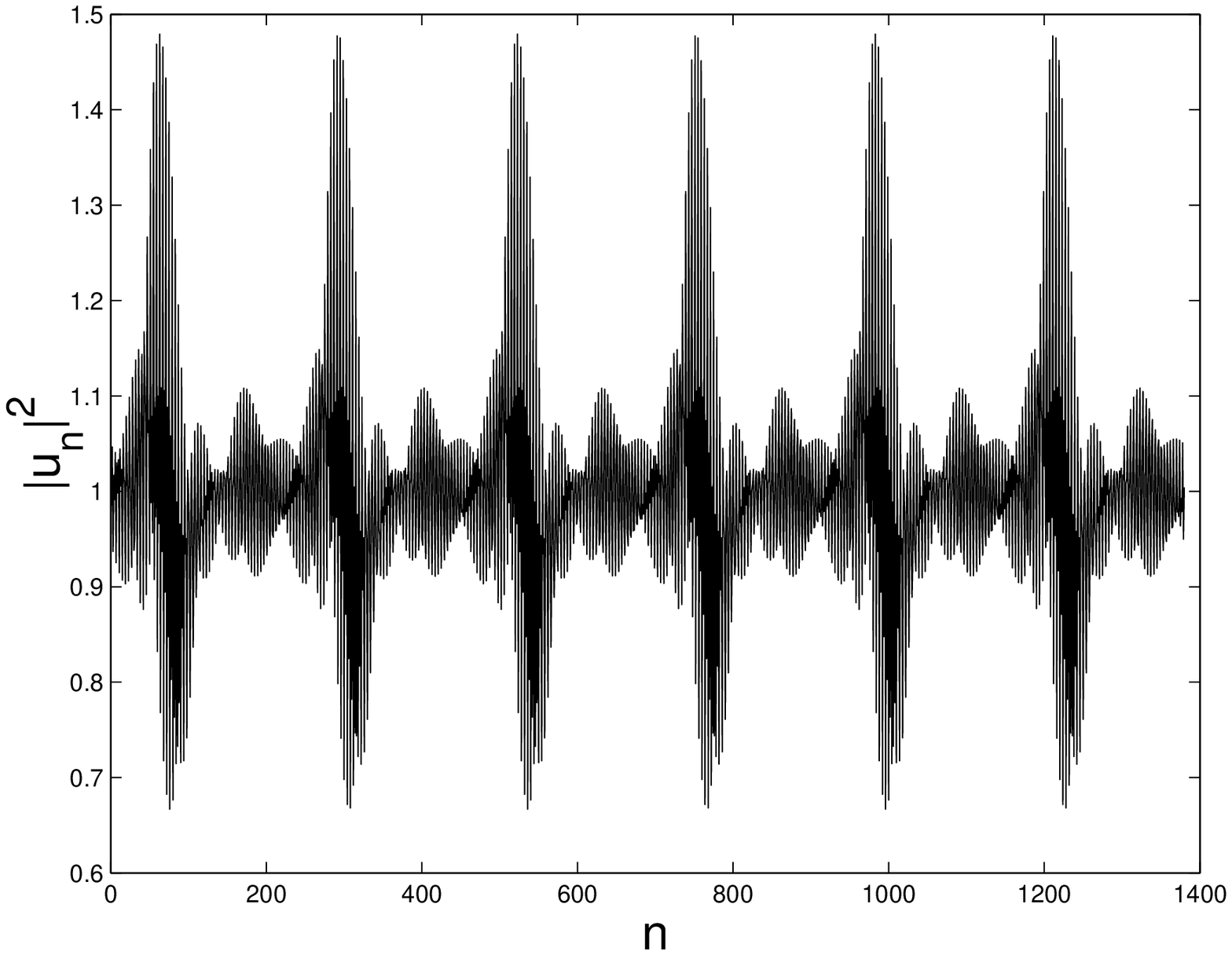, width=6.3cm,angle=0, clip=}}
\caption{The case which is modulationally 
unstable. The left panel shows the square modulus of the field, before 
the onset of the instability at $t=17$, and  right  after the onset, 
at $t=56$. In this case $q=1.58446$.}
\label{dfig3}
\end{figure}

Finally, in Fig. \ref{dfig3} we show the time evolution of a modulationally
unstable case for $q=1.58446$. One clear feature that distinguishes 
the parametric
instability shown in Fig. \ref{dfig2} from the modulational one of Fig.
\ref{dfig3} is the time scale of the instability development, or 
equivalently the corresponding instability growth rate. 
The latter appears to be
much larger in the case of the modulational instability; as a result, the
MI sets in much sooner for equal magnitude perturbations of the two plane
waves.

From the growth rate predictions of Eq. $10$ of \cite{kivshar92} and
from our analysis,
it can be calculated (in the cases shown) 
that the theoretical growth rate in the case of the 
modulational instability is approximately $0.23218$ (for$q=1.58446$)
and the corresponding  growth rate in the case of the parametric 
instability is approximately $0.00115$ (for $q=1.51161$).
The considerably larger growth rate of the MI qualitatively justifies the
earlier temporal development of the latter.

\section{Conclusions}

In this short communication, we have examined the potential
of time-dependent coefficients, in the context of a non-autonomous
discrete nonlinear Schr{\"o}dinger equation, to induce a parametric
instability of plane wave states. Such an instability has been 
identified and its boundaries established at the level of a leading
order theory within a multiscale expansion. The theoretical findings
have been numerically tested through direct simulations and have been
found to be in agreement with the theoretical predictions (except
for the regime between the parametric and modulationally unstable wavenumbers,
where higher order parametric resonances may ensue).

The distinctive feature of the present parametric instability is that
it has a much longer threshold time for its dynamical development 
in comparison with the modulational instability. This can be both
quantified by means of the comparison of their respective growth 
rates, as well as observed in the course of direct numerical 
simulations. Studying the effects of parametric instabilities
in other contexts including that of coherent, nonlinear wave structures
in BEC is very relevant and will be addressed in future publications.

PGK gratefully acknowledges support from NSF-DMS-0204585 and from
the Eppley Foundation for Research.

\end{document}